\documentclass[letterpaper, 10 pt, conference]{ieeeconf}  

\usepackage{xcolor}
\usepackage{amsfonts}
\usepackage{amsmath}
\usepackage{graphicx}
\usepackage{caption}
\usepackage{subcaption}
\usepackage{cleveref}
\usepackage{bbm}
\usepackage{url}

\usepackage{tikz}
\newcommand\copyrighttext{%
  \footnotesize \textcopyright \the\year{} IEEE. Personal use of this material is permitted.  Permission from IEEE must be obtained for all other uses, in any current or future media, including reprinting/republishing this material for advertising or promotional purposes, creating new collective works, for resale or redistribution to servers or lists, or reuse of any copyrighted component of this work in other works.}

\newcommand\copyrightnotice{%
\begin{tikzpicture}[remember picture,overlay]
\node[anchor=south,yshift=10pt] at (current page.south) {\fbox{\parbox{\dimexpr0.75\textwidth-\fboxsep-\fboxrule\relax}{\copyrighttext}}};
\end{tikzpicture}%
}

\newcommand{\ch}[1]{\textcolor{black}{#1}}

\usepackage{soul}  

\IEEEoverridecommandlockouts                              

\title{\LARGE \bf
A Data-Informed Analysis of Scalable Supervision for Safety in Autonomous Vehicle Fleets
}

\author{Cameron Hickert$^{1}$, Zhongxia Yan$^{2}$, and Cathy Wu$^{3}$
\thanks{*This work was partially supported by the MIT Amazon Science Hub, MIT-IBM Watson AI Lab, MIT Energy Initiative (MITEI) Mobility Systems Center, MIT Mobility Initiative, and National Science Foundation (NSF) under grant number 2149548.}
\thanks{$^{1}$Cameron Hickert is with the Institute for Data, Systems, and Society, Massachusetts Institute of Technology,
        Cambridge, MA 02139, USA.
        {\tt\small chickert@mit.edu}}%
\thanks{$^{2}$Zhongxia Yan is with the Department of Electrical Engineering and Computer Science, Massachusetts Institute of Technology,
        Cambridge, MA 02139, USA.
        {\tt\small zxyan@mit.edu}}%
\thanks{$^{3}$Cathy Wu is with the Laboratory for Information \& Decision Systems; the Institute for Data, Systems, and Society; and the Department of Civil and Environmental Engineering, Massachusetts Institute of Technology,
        Cambridge, MA 02139, USA.
        {\tt\small cathywu@mit.edu}}%
}

\begin{document}

\maketitle
\copyrightnotice
\thispagestyle{empty}
\pagestyle{empty}

\begin{abstract}

Autonomous driving is a highly anticipated approach toward eliminating roadway fatalities. At the same time, the bar for safety is both high and costly to verify.
This work considers the role of remotely-located human operators supervising a fleet of autonomous vehicles (AVs) for safety. 
Such a `scalable supervision' concept was previously proposed to bridge the gap between still-maturing autonomy technology and the pressure to begin commercial offerings of autonomous driving.
The present article proposes DISCES, a framework for \ch{Data-Informed Safety-Critical Event Simulation}, to investigate the practicality of this concept from a dynamic network loading standpoint. 
With a focus on the safety-critical context of AVs merging into mixed-autonomy traffic, vehicular arrival processes at 1,097 highway merge points are modeled using microscopic traffic reconstruction with historical data from interstates across three California counties.
Combined with a queuing theoretic model, these results characterize the dynamic supervision requirements and thereby scalability of the teleoperation approach. Across all scenarios we find reductions in operator requirements greater than 99\% as compared to in-vehicle supervisors for the time period analyzed.
The work also demonstrates two methods for reducing these empirical supervision requirements: (i) the use of cooperative connected AVs --- which are shown to produce an average 3.67 orders-of-magnitude system reliability improvement across the scenarios studied --- and (ii) aggregation across larger regions.

\end{abstract}

\section{Introduction}

Autonomous vehicle (AV) deployments in the real world face a number of obstacles, suggesting that `Level 5' autonomy --- in which AVs can drive safely and effectively at all times, in all places, and in all conditions --- is further away than initially projected~\cite{bansal2017forecasting}.
In the near- to medium-term, and longer if necessary, vehicles with some degree of autonomous capabilities still rely on human drivers for supervision and operation in various driving situations. 
Should a Level 5 AV be developed, the question of how to prove its safety likewise remains open, given the unpredictable and rare nature of many dangerous driving circumstances~\cite{kalra2016driving}. 

At the same time, AVs have the potential to significantly improve system-level roadway performance and reduce carbon emissions, even when they only represent a minority of vehicles in a traffic system~\cite{wu2021flow}.

In light of these realities, AV deployments today make use of remote operators: Cruise on average triggered one remote intervention every 2.5 to 5 miles in its San Francisco AV deployment~\cite{mickle2023gm}.
A number of companies are directly targeting `teledriving' as their primary approach to addressing autonomous technology's shortcomings~\cite{amador2022survey}.

While previous research has considered how to scale online human supervision for a desired level of safety for merging AVs~\cite{hickert2023cooperation}, it performs queueing-theoretic statistical analyses of various scenarios rather than considering real-world traffic data. 

Focusing on the case of AVs merging into mixed-autonomy traffic, our work takes a data-informed dynamic network loading approach. By leveraging large-scale reconstructions of historical data, we simulate traffic across the course of a day on 1,097 freeway merge points in Los Angeles, San Bernardino, and Orange Counties to empirically assess the question: how many supervisors would we need for county-level human oversight of dangerous interstate merges over an entire day?

More specifically, the work makes three contributions:
\begin{enumerate}
    \item The introduction of DISCES: a framework for \ch{Data-Informed Safety-Critical Event Simulation}. \ch{This combines large-scale data with traffic microsimulation to approximate the number and location of critical safety events in realistic settings}.
    \item Empirical estimates \ch{for the number of human operators needed to supervise AV merges on county-scale interstate networks, as well as queuing theoretic estimates for long-term supervision needs} in these settings. 
    \item Demonstration of two methods for \ch{improving supervision scalability: cooperative connected AVs and supervision task aggregation} across larger regions.
\end{enumerate}

Our findings provide data-driven validation for the idea that human supervision may ease deployment of imperfect AVs in certain safety-critical roadway settings.

\section{Related Work}

\paragraph{Data-informed analysis of traffic safety} Assessing safety in traffic systems historically has been driven by observational studies using historical datasets, including police-reported crash data~\cite{arun2021systematic}. As a result, works assessing transportation system safety have often leaned on implicit notions of safety as the absence of crashes, injuries, or deaths for analysis~\cite{duduta2015traffic, sakhapov2018traffic}, even when considering broader public health and risk management frameworks for addressing traffic safety issues~\cite{ederer2023safe}. Of course, drivers know anecdotally that many unsafe situations (‘close calls’) do not result in accidents, and even fewer may result in reported accidents. This is where traffic modeling is of use.

Traffic modeling is typically conducted at one of two levels: macroscopic (aggregated) and microscopic (agent-based). Macroscopic traffic system models generally do not account for safety incidents~\cite{papageorgiou1998some}, although there is some discussion of safe velocities in traffic flows~\cite{khan2021macroscopic}. This is partially due to the fact that safety incidents are microscopic in nature (that is, partially dependent on local interactions) and macroscopic models do not capture such vehicle-level dynamics. 

Traffic microsimulators such as the Simulation of Urban Mobility (SUMO) are therefore well-suited to safety analysis given their modeling of realistic vehicle-level traffic dynamics~\cite{lopez2018microscopic}.
However, to the authors' knowledge, such microsimulation tools have not been paired with real-world data for safety analysis on the scale of this work. Work combining observational data and microsimulation has been small-scale and focused on assessing various car-following models~\cite{guido2011safety}. Larger-scale works leveraging traffic simulators focus on performance metrics like throughput or travel time, but ignore safety metrics in their analysis~\cite{kouvelas2011adaptive, zlatkovic2019assessment}. The traffic safety survey in \cite{du2023safety} only relies on historical data for safety assessments, whereas microsimulation models provide an opportunity to analyze counterfactuals.


\paragraph{Teleoperation for Connected Autonomous Vehicles (CAVs)}

Teleoperation of roadway vehicles (also known as `teledriving,' `tele-assistance,' or remote operation) is growing in popularity alongside advances in cellular network infrastructure and vehicular connectivity and autonomy \ch{--- o}ne recent survey found 15 companies offering operations support or vehicle service for remote driving, assistance, or monitoring~\cite{amador2022survey}. Some of these, such as Cruise, combine remote human operators with CAV technology~\cite{mickle2023gm}. Such deployments exhibit varying objectives including performance, customer satisfaction, and safety.

Academic literature on the subject tends to focus on human factors questions, such as teleoperation interface design~\cite{kuru2021conceptualisation, tener2022driving, tener2023toward}. A small but growing body of work applies queuing models to assess the scalability of teleoperation for connected vehicle fleets~\cite{hickert2023cooperation, benjaafar2022human}.

To the authors' knowledge, no work exists leveraging historical data to assess the scalability of such teleoperation, especially for AV safety.

\paragraph{Highway merging}
This paper's focus is merging, a known challenging situation for AVs~\cite{jula2000collision, zhou2016impact}. It often occurs at high speeds with the potential for collision with other vehicles or fixed infrastructure and it depends on the behavior of nearby vehicles, perhaps even necessitating jockeying or (for humans) hand signaling that evades easy explicit definition or even violates standard driving rules. Indeed, previous work modeling safety at freeway merges found lower speeds to be associated with \textit{more} collisions on inbound lanes~\cite{ahammed2008modeling}. Due to the challenging and dangerous nature of merges, we focus on this maneuver, but future work could consider supervision events beyond merges (lane changes, handling construction zones, etc.) using the DISCES framework.

\section{Preliminaries}

\paragraph{Reachability}

Reachability analysis is a well-known tool for assessing and enforcing system safety in robotics. Increasingly it has been applied to AVs~\cite{althoff2010reachability}. It has a straightforward application in our context: two vehicles may collide in some time horizon only if they can reach each other in that horizon. Thus, if the vehicles are beyond each other's reachable zone, they cannot collide. And where vehicles cannot collide, from a merging vehicle's perspective, the merge maneuver is akin to driving straight on an open lane. 
Previous work investigating the scalability of supervision in this context leveraged kinematics-based reachability to determine the scenarios which require human supervision~\cite{hickert2023cooperation}.

This work adopts the same approach: a merging AV must be supervised if there is any possibility over some time horizon $h$ of collision with a vehicle on the lane to which it is merging.
Such a method is an over-approximation of the reachable set, but it maintains conservatism and is computationally efficient.
Additionally, by \textit{not} supervising merges where an AV \textit{cannot} collide into a vehicle on the lane to which it is merging, we can achieve better supervision scalability.
Note that this assumes a baseline level of AV capabilities (e.g., \ch{lane-keeping}). 

\paragraph{Queuing theoretic model}

Our simulation \ch{allows us to find the number of human supervisors that would have been necessary during the period simulated}. Of course, given that this is a hindsight estimate, a reasonable question is: how can we size supervision teams appropriately, given uncertainty in traffic flows? 

To address this, we leverage a result from previous work indicating the ``fraction of AVs that require supervision but cannot immediately receive it (and thus go unsupervised)''~\cite{hickert2023cooperation} under independent arrival and service rate assumptions. This is given by
\begin{equation}
\label{eq:P_k}
    \begin{split}
        P_k = \frac{(\lambda/\mu)^k / k!}{\sum\limits_{i=0}^{k} (\lambda / \mu)^i / i!},
    \end{split}
\end{equation}

where the arrival of merges to supervise arise via a Poisson process $\text{Poisson}(\lambda)$, $k$ is the number of supervisors for whom the service time of each follows $\text{Exp}(\mu)$, and $\lambda < \mu$ to have a steady state probability. This allows us to model the number of supervisors necessary to achieve an arbitrary reliability level $1 - P_k$ (e.g., 99\%, 99.99\%) over time.

Since previous work did not have data-driven values for $\lambda$ and $\mu$, we generate these values via traffic reconstruction and leverage them to identify the empirical number of supervisors needed to achieve a given reliability level at the rates found for the given day, as well as to demonstrate the benefits of pooling supervision tasks.

\section{Problem Formulation}
\label{sec:problem_formulation}

We are interested in (i) the number of operators $\hat{k} \in \mathbb{N}$ necessary to supervise all safety-critical events over the course of a day, and (ii) the minimum number of operators $k^* \in \mathbb{N}$ necessary to achieve a long-term desired level of supervision reliability $1 - \epsilon$ such that $P_{k^*} < \epsilon$ based on the queuing theoretic model in~\Cref{eq:P_k}, for which we will determine $\lambda$ and $\mu$ empirically. 

Let $A_{j,m,t}$ indicate the event in which a merging AV $j$ can arrive at merge point $m$ within the near-future of time $t$. Let $A_{i,m,t}$ be defined similarly, except for an on-highway vehicle $i$ --- either human-driven or an AV --- on the lane to which AV $j$ is about to merge.

We can thus let 
\begin{equation} \label{eq:conflict}
    C_{j,m,t} = \mathbb{I}\{ \exists i A_{i,m,t} \wedge A_{j,m,t} \}
\end{equation}
indicate a \textit{potential} near-term conflict between on-highway vehicle $i$ and merging AV $j$ at merge point $m$ and time $t$, where \ch{$\mathbb{I}$ is an indicator function}. This is the safety-critical event that \ch{we require a} human operator to supervise. Therefore, the number of AVs requiring supervision at a given time $t$ is $s_t = \sum_{j=0}^{n} C_{j,m,t}$, where $n$ is the number of AVs in the system at that time. 

For the case in (i) \ch{above}, we can find the number of supervisors needed over the course of the time period with a length of $T$ as 
\begin{equation} \label{eq:max_num_supervisors}
    \hat{k} = \max_{t \in \{1, ... , T\}}s_t.
\end{equation}

For the case in (ii) \ch{above}, we seek to extract empirical values of $\hat{\lambda}$ and $\hat{\mu}$ from the data-driven simulation and then leverage~\Cref{eq:P_k} to find the number of supervisors $k^*$ to achieve an arbitrary desired reliability level $1 - \epsilon$ such that $P_{k^*} < \epsilon$. 

\section{Methodology}

We term the framework for our approach DISCES, for \ch{Data-Informed Safety-Critical Event Simulation}. It consists of four parts:

\begin{enumerate}
    \item Large-scale, real-world traffic data
    \item Traffic reconstruction via microsimulation based on that data
    \item Extraction of safety-critical events
    \item Analysis of the extracted safety-related events
\end{enumerate}

In short, data is used for dynamic network loading to calibrate a traffic reconstruction, which is in turn used to simulate the dynamics necessary to identify and extract safety-critical events at the level of individual vehicles. These events are then passed to the \ch{fourth} module for handling and analysis. 

\paragraph{Data-driven traffic reconstruction --- (1) and (2) above}

DISCES relates to the broader class of approaches to the dynamic network loading challenge --- namely, determining time-varying traffic flows across a road network. Given that vehicle detection data from stationary sensors is not sufficient to fully determine specific routes and traffic volume, the first step is to reconstruct realistic, data-informed traffic flows.

As described in~\cite{qu4687440revisiting}, a road network $G = (V, E)$ consists of edges (road segments) $E$ and vertices $V$. We have real-world vehicle detection count data $c_t(e)$ at corresponding edges $e \in E$ over time $0 \leq t \leq T$. Our objective is to recover a set of routes $R$ and a flow function $g_t$ mapping a route $r \in R$ of edges to the number of vehicles on each edge in $r$ at time $t$. To do so, one must find $R$ and $g_t$ such that, $\forall e \in E$, $ c_t(e) \approx \sum_{r \in R}g_t(r, e)$, where $g_t(r, e)$ is the element of $g_t(r)$ corresponding to $e$. This \ch{can} be accomplished via a variety of methods; we describe ours in the next section.

Vehicle inflows $g_t(r, r_0)$, where $r_0$ is the first edge along $r$, are used to induce traffic system states $x_t$ in a traffic dynamics model $F$ (e.g., traffic microsimulator SUMO) which simulates the next state $x_{t + 1} = F(x_t, u_t, g_t)$ given the current state and control inputs $u_t$. Traffic microsimulators are known as one approach towards dynamic network loading, particularly with the growing availability of vehicle detection data~\cite{tsanakas2021data}. The output is thus a data-driven traffic reconstruction from which we can extract safety-critical events. It is important to note that --- due to coarseness in data granularity, underspecification of routes, \ch{AV-induced trajectory shifts,} etc. --- we do not expect the simulation to be a perfect recreation. Fortunately, we are not interested in recreating \textit{specific} safety-critical events (e.g., a particular crash on August 1st), but rather extracting system-level safety statistics (e.g., the number of supervisors that we would need on days \textit{like} August 1st). Still, as the quality and quantity of traffic data grows --- for example, from mobile devices or onboard navigation systems --- so can the quality of the reconstruction. 

\begin{figure*}
  \centering
  \begin{subfigure}[b]{0.26\textwidth}
    \includegraphics[width=\textwidth]{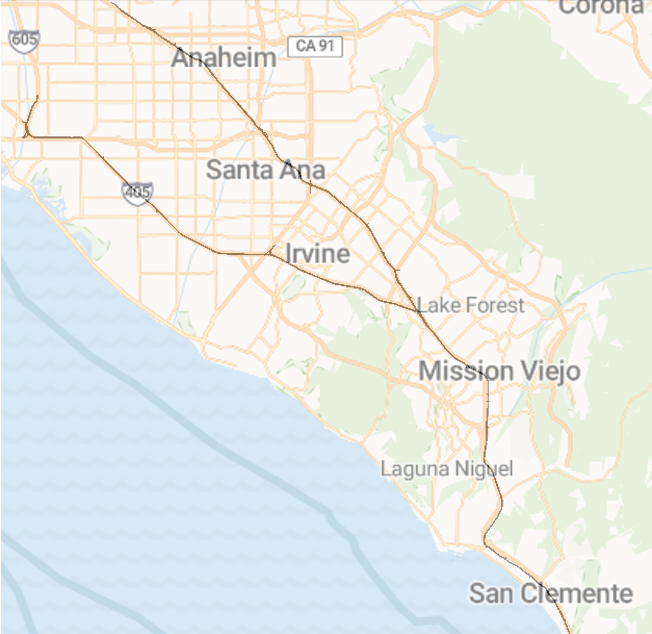}
    \caption{Orange County}
    \label{fig:orange_network}
  \end{subfigure}
  \hspace{0.1cm}
  \begin{subfigure}[b]{0.26\textwidth}
    \includegraphics[width=\textwidth]{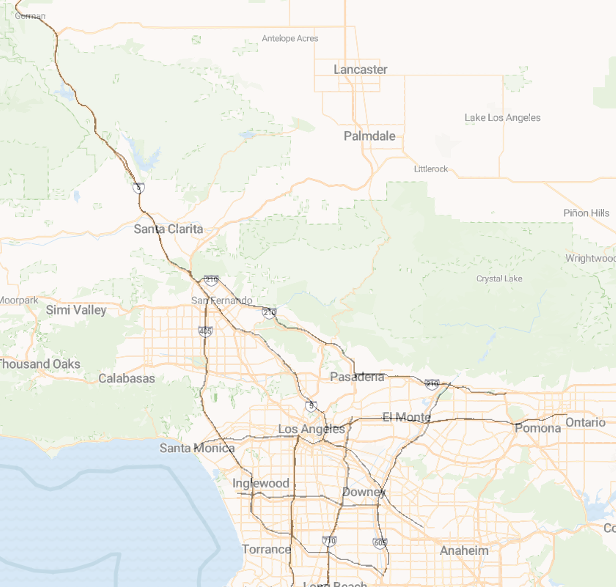}
    \caption{Los Angeles County}
    \label{fig:la_network}
  \end{subfigure}
  \hspace{0.1cm}
  \begin{subfigure}[b]{0.43\textwidth}
    \includegraphics[width=\textwidth]{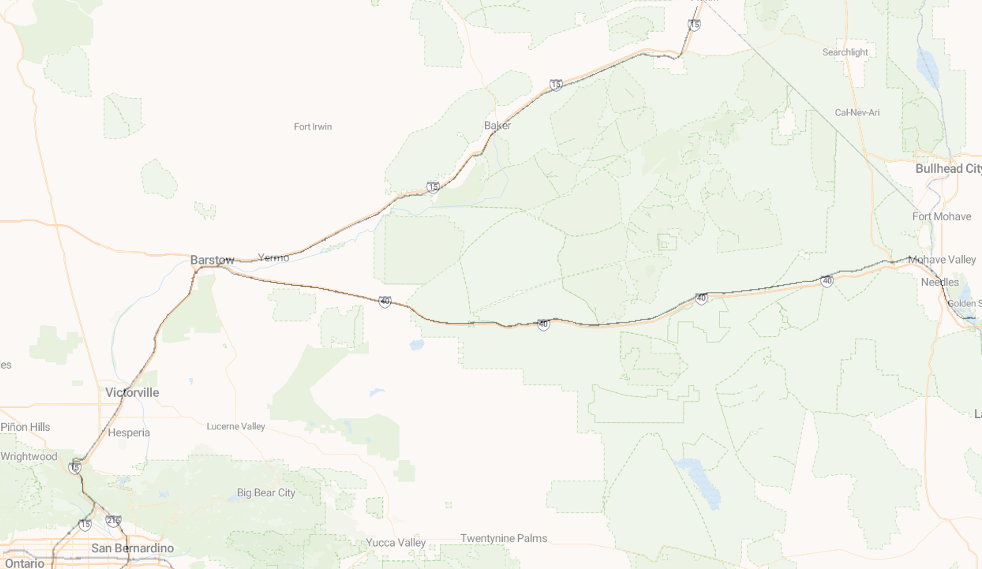}
    \caption{San Bernardino County}
    \label{fig:sb_network}
  \end{subfigure}
  \caption{Visualizations of each county and its OpenStreetMap interstate network overlaid in black.\protect\footnotemark}
  \label{fig:osm_nets}
\end{figure*}

\paragraph{Extraction and analysis of safety-critical events --- (3) and (4) above} From the traffic dynamics model $F$, at each time $t$ the number of safety-critical events $s_t$ \ch{can} be extracted from the system state $x_t$. This \ch{can} be done via a function $z$ such that $s_t = z(x_t)$. Importantly, the DISCES framework is suitable for a range of safety assessment functions $z$. Below we present a reachability-based safety event identification, but one could use any number of safety surrogates identifiable in simulation, whether based on time, deceleration, distance, kinetic energy, or some combination of these.

A conservative estimate of the reachable distance over time horizon $h$ for a vehicle $i$ can be written via kinematics as 
\begin{equation} \label{eq:d_kinematics}
    d_{i}(t, h) = v_{i,t} h + \frac{1}{2} a_{i, max} h^2,
\end{equation}
where $v_{i,t}$ is the vehicle's speed at current time $t$ and $a_{i, max}$ is the maximum acceleration. Indicating the distance between a vehicle $i$ and merge point $m$ at time $t$ as $d_{i, m}(t)$, we can rewrite~\Cref{eq:conflict} using reachability principles as
\begin{equation*}
    C_{j,m,t} = \mathbb{I}\left\{\exists i \left(d_{i}(t, h) \geq d_{i,m}(t)\right) \wedge \left(d_{j}(t, h) \geq d_{j,m}(t)\right)\right\}.
\end{equation*}
This represents the event in which the merge point falls within the reachable distance of both a merging AV $j$ and at least one on-highway vehicle $i$. 

Thus we can find the value of $s_t$ and $\hat{k}$ for case (i) as described above, \ch{for example} using $T =$ 86,400 in~\Cref{eq:max_num_supervisors} for a daylong simulation with seconds-level granularity. This provides the number of supervisors needed over the day.

For case (ii) recall we aim to extract empirical values of $\lambda$ and $\mu$ for use in~\Cref{eq:P_k}. The number of supervision tasks for the $l$th hour, $L = \{l * 3600, ..., (l+1) * 3600\}$, in a simulation with seconds-level time granularity is
\begin{equation*}
    q_l = \sum_{j=0}^{n} \sum_{m \in M(j)} \mathbb I\left\{C_{j,m,t} = 1\ \exists t \in L\right\},
\end{equation*}
where $M(j)$ is the set of merge points $j$ encounters, such that a conflict event only registers once per merge point per AV (that is, per $j, m$ pair). Intuitively, once a remote operator begins supervising the merge, they continue to do so until the merge is complete or supervision is unnecessary for the remainder of the merge, so there is no need to re-engage each timestep. Note this formulation allows the possibility for an AV to encounter multiple merge points on its journey --- indeed, in the simulation many do.

\footnotetext{Modified map tiles from Stamen Design, under CC BY 4.0. Data by OpenStreetMap, under ODbL.}
\addtocounter{footnote}{1}
\footnotetext{Image modified from public domain source. Original image by Thadius Miller, \url{https://commons.wikimedia.org/wiki/File:California_county_map_(labeled).svg}, 06/23/2010.}

Thus, the average arrival rate of supervision tasks for the $l$th hour with seconds-level granularity is
\begin{equation} \label{eq:lambda}
    \hat{\lambda_l} = \frac{q_l}{3600}.
\end{equation}

Similarly, the value parameterizing the service rates of the supervisors for the $l$th hour in the same simulation is: 
\begin{equation} \label{eq:mu}
    \hat{\mu_l} = \frac{b_l}{q_l},
\end{equation}
where
\begin{equation*}
    b_l = \sum_{j=0}^{n} \sum_{m \in M(j)} \sum_{t \in L} C_{j,m,t}
\end{equation*}
represents the cumulative supervision time required. In short, $\hat{\mu_l}$ is the average time per supervision task in the $l$th hour, again allowing for an AV to encounter multiple merge points.

\section{Experimental Setup}

\addtocounter{footnote}{-1}

\begin{figure}
    \centering
    \includegraphics[width=0.4\textwidth]{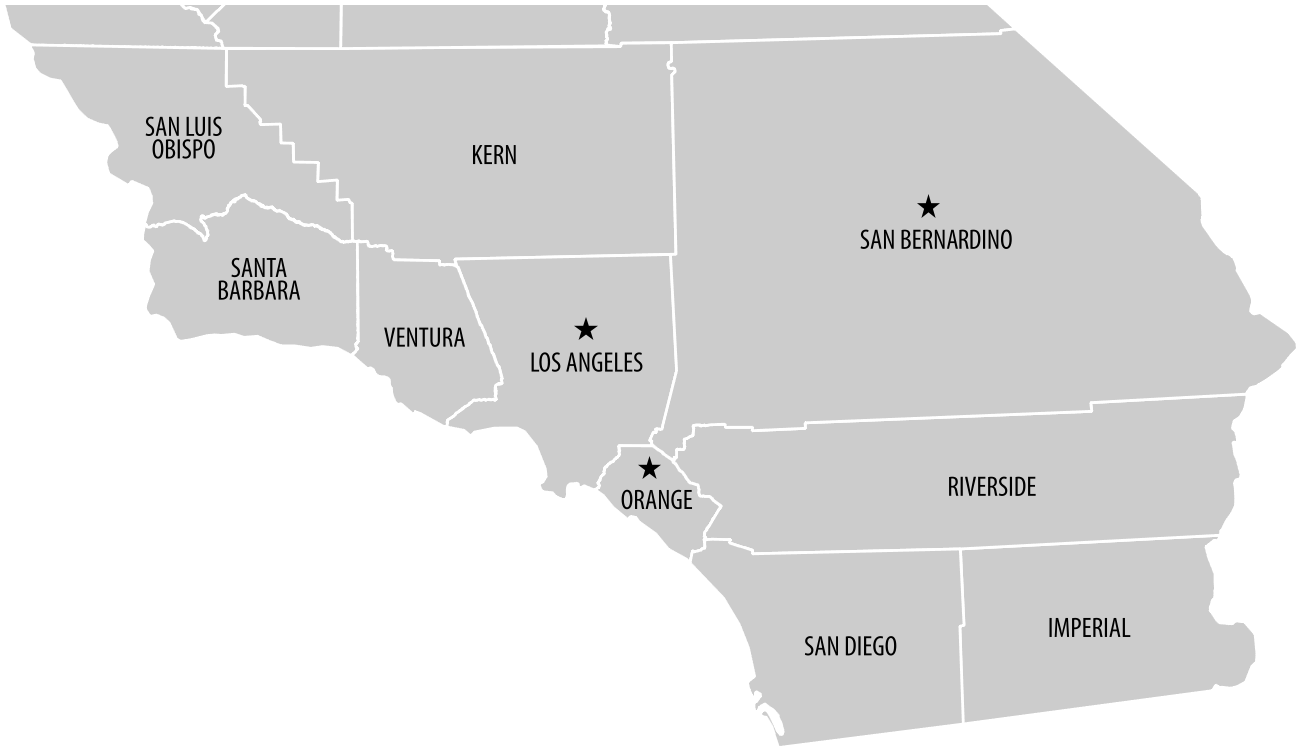}
    \caption{A map showing counties in southern California. The three selected counties are indicated with stars. Their diverse geographic areas and populations make them interesting study cases, and their proximity to each other suits the supervisor aggregation analysis.\protect\footnotemark}
    \label{fig:socal}
\end{figure}

\begin{figure}
    \centering
    \includegraphics[width=0.45\textwidth]{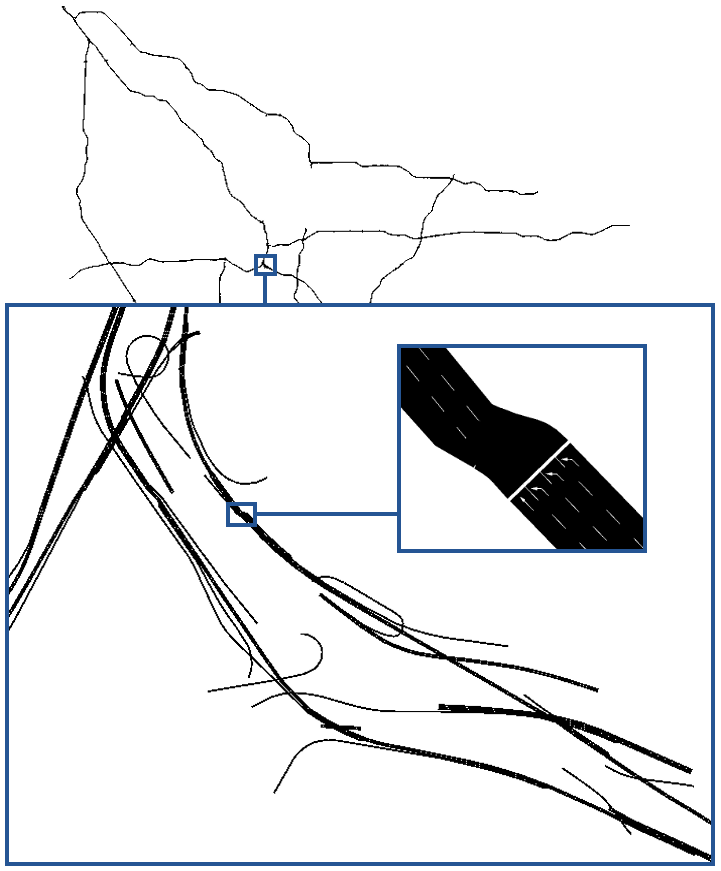}
    \caption{As lane-level detail cannot be seen in \Cref{fig:osm_nets}, this figure shows greater detail for a subsection of the Los Angeles County traffic network reconstruction, as well as a lane-level pop-out.}
    \label{fig:lane-popout}
\end{figure}

\paragraph{Traffic microsimulation with realistic traffic data}

For the first and second subcomponents of the DISCES framework, we adapted the traffic flow reconstruction and simulation outlined in~\cite{qu4687440revisiting}, with two notable modifications. First, we use 1-second time discretization instead of 5-second time discretization for increased \ch{simulation granularity and to enable seconds-level reachability analysis}. Second, we use Intelligent Driver Model (IDM) values to parameterize the driving behavior of the vehicles in the simulation~\cite{treiber2013traffic}. 
These are drawn from German highway drivers; we found they provided similar qualitative behavior to values drawn from US highway drivers, but are more interpretable and were found to create fewer simulation artifacts than those in~\cite{qu4687440revisiting}. Most importantly, IDM values are significantly more common in assessments of safe driving behavior, and thus we focus our reporting and discussion on simulations based upon them. 

This pipeline produces vehicle flows approximating historical traffic via SUMO's in-built method for computing vehicle routes from vehicle detection data, based on a maximum flow algorithm suggested in~\cite{behrisch2018route}. For the road network $G$, OpenStreetMaps data is used to build interstate networks at lane-level granularity, including on- and off-ramps~\cite{OpenStreetMap}. California's Caltrans Performance Measurement System (PeMS), which has vehicle detector data in 30-second increments from nearly 40,000 sensors across the state's freeway system, is the source of vehicle count data $c_t(e)$~\cite{Caltrans2023}.

We simulate interstate traffic across three adjacent California counties (Los Angeles, Orange, and San Bernardino) based on the historical flows for the 24 hours of Wednesday, August 1st, 2018 in those counties. \ch{This was selected as one example of a standard weekday --- future work could consider days with atypical travel demand.} See Figures 1, 2, and 3. These were chosen for their significant traffic volumes, variety in density and geography, and because their proximity could allow for easier pooling of supervisors than non-adjacent counties \ch{(due to teleoperator location, system latency issues, consistency of local driving laws and norms, similarity of driving conditions, etc.)}. A merge is defined as any SUMO junction that has fewer outgoing lanes than incoming lanes. This includes on-ramps where a merge is necessary, as well as regular interstate portions where the number of lanes is reduced. This definition identified 192 merge points in the Orange County interstate system, 289 in San Bernardino County, and 616 in Los Angeles County.

This work considers three types of AVs: unconnected (UCAVs), noncooperative connected (NCAVs), and cooperative connected (CCAVs). 
\ch{The first difference is that UCAVs and NCAVs do not adjust their behavior to accommodate merging AVs, while CCAVs do. That is, a}ll AVs are parameterized with the same driving models as the human vehicles (HVs), except the CCAVs, which have an additional simple cooperative driving policy whereby they seek to \ch{free a lane for a merging AV by shifting to an adjacent lane when space allows}, similar to the altruistic behavior described in~\cite{li2021employing}.
\ch{The second difference is that, unlike the UCAVs, t}o emulate the advantages of connectivity for reducing uncertainty, the NCAVs and CCAVs have truncated reachable zones as in~\cite{hickert2023cooperation}. That is, since the vehicles are autonomous and connected, they communicate their near-term trajectory to other connected AVs. This reduces the uncertainty inherent to the calculation of reachable sets and truncates the kinematics-based reachable distance in~\Cref{eq:d_kinematics} to the vehicle's length.

This work takes $h=5$ seconds for our reachability time horizon. One meta-analysis of 129 studies found more than 80\% of the 520 mean times for human takeover of vehicular control from an automated system were below five seconds, even while several of the included studies assessed distracted drivers~\cite{zhang2019determinants}. Additional research including participants driving at highway speeds also exhibit takeover times below five seconds~\cite{eriksson2017takeover}. Recent AV deployments suggest this threshold may be an over-approximation: 98\% of Cruise's remote assistance sessions for its San Francisco deployment were answered within three seconds~\cite{kolodny2023cruise}. Finally, note that in practice, vehicles rarely drive their maximum acceleration, so the $h=5$ reachability horizon actually provides considerably more time for a supervisor to assume control of the AV before it reaches the merge point. 

We simulate nine scenarios for each county, representing the possible combinations of three AV penetration rates (25\%, 50\%, and 75\%) for our three AV types. In the case that a merging AV has on-and-off supervision requirements (e.g., needs a supervisor, then does not because the on-highway vehicle shifted lanes, then does when another on-highway vehicle approaches), we take the conservative approach by assigning a supervisor for the duration from the first moment of required oversight to the last. 

\section{Results}

\begin{table*}
\centering 
\begin{tabular}{|c|c|cccccc|}
\hline
 &  & \multicolumn{2}{c|}{\textbf{25\% AV Pen. Rate}} & \multicolumn{2}{c|}{\textbf{50\% AV Pen. Rate}} & \multicolumn{2}{c|}{\textbf{75\% AV Pen. Rate}} \\ \cline{3-8}
\textbf{County} & \textbf{AV Type} & Mean (Std.)  & Max. [Mean] & Mean (Std.) & Max. [Mean] & Mean (Std.) & Max. [Mean] \\
&  & Speed (m/s) & \# Suprvsrs. & Speed (m/s) &\# Suprvsrs. & Speed (m/s) & \# Suprvsrs. \\ \hline
& UCAV &   18.83 (4.04) &        24 [7.2] &   18.83 (4.04) &       45 [14.4] &   18.83 (4.04) &       59 [21.5] \\ \cline{2-8}
\textbf{Orange} & NCAV & 18.83 (4.04) &        \textbf{22 [6.3]} &   18.83 (4.04) &       37 [10.5] &   18.83 (4.04) &       41 [12.4] \\ \cline{2-8}
& CCAV & 18.83 (4.03) &        24 [5.8] &    18.8 (4.03) &        \textbf{30 [9.6]} &   18.74 (4.05) &       \textbf{37 [10.8]} \\ \cline{2-8}
3.1mn pop. & Baseline 1: all merges & - &         42 [12] &              - &         75 [24] &              - &        101 [36] \\ \cline{2-8}
800 mi$^2$ area& Baseline 2: trip duration & - &     3,406 [1,861] &              - &     6,785 [3,727] &              - &    10,025 [5,571]  \\ \cline{2-8}
192 merge points& CCAV-UCAV gain & 0.00\% &            0.0\% &         -0.16\% &           33.3\% &         -0.48\% &           37.3\% \\ \cline{2-8}
& CCAV-Baseline 1 gain & - &           42.9\% &              - &           60.0\% &              - &           63.4\% \\ \cline{2-8}
& CCAV-Baseline 2 gain & - &           99.3\% &              - &           99.6\% &              - &           99.6\% \\ \hline
& UCAV & 23.98 (2.04) &        22 [5.7] &   23.98 (2.04) &       35 [11.4] &   23.98 (2.04) &       49 [17.0] \\ \cline{2-8}
\textbf{San} & NCAV & 23.98 (2.04) &        \textbf{20 [4.8]} &   23.98 (2.04) &        \textbf{27 [7.3]} &   23.98 (2.04) &        \textbf{30 [7.5]} \\ \cline{2-8}
\textbf{Bernardino} & CCAV & 23.88 (2.09) &        \textbf{20 [4.9]} &    23.6 (2.24) &        33 [7.3] &   23.23 (2.36) &        \textbf{30 [7.7]} \\ \cline{2-8}
& Baseline 1: all merges & - &         46 [15] &              - &         77 [28] &              - &        111 [41] \\ \cline{2-8}
2.1mn pop. & Baseline 2: trip duration & - &     3,400 [1,686] &              - &     6,813 [3,387] &              - &    10,171 [5,074]  \\ \cline{2-8}
20,000 mi$^2$ area & CCAV-UCAV gain & -0.42\% &            9.1\% &         -1.58\% &            5.7\% &         -3.13\% &           38.8\%  \\ \cline{2-8}
289 merge points & CCAV-Baseline 1 gain & - &           56.5\% &              - &           57.1\% &              - &           73.0\% \\ \cline{2-8}
& CCAV-Baseline 2 gain & - &           99.4\% &              - &           99.5\% &              - &           99.7\% \\ \hline
& UCAV & 22.24 (2.53) &        32 [8.1] &   22.24 (2.53) &       48 [16.2] &   22.24 (2.53) &       67 [24.2] \\ \cline{2-8}
\textbf{Los} & NCAV & 22.24 (2.53) &        29 [7.0] &   22.24 (2.53) &       41 [11.5] &   22.24 (2.53) &       49 [13.0] \\ \cline{2-8}
\textbf{Angeles}& CCAV & 22.22 (2.55) &        \textbf{25 [7.0]} &   22.19 (2.55) &       \textbf{36 [11.3]} &   22.13 (2.57) &       \textbf{38 [12.0]} \\ \cline{2-8}
 & Baseline 1: all merges & - &         38 [10] &              - &         60 [20] &              - &         82 [30] \\ \cline{2-8}
9.7mn pop. & Baseline 2: trip duration & - &     5,067 [2,653] &              - &    10,181 [5,317] &              - &    15,329 [7,953] \\ \cline{2-8}
4,000 mi$^2$ area & CCAV-UCAV gain & -0.09\% &           21.9\% &         -0.22\% &           25.0\% &         -0.49\% &           43.3\% \\ \cline{2-8}
616 merge points & CCAV-Baseline 1 gain & - &           34.2\% &              - &           40.0\% &              - &           53.7\% \\ \cline{2-8}
& CCAV-Baseline 2 gain & - &           99.5\% &              - &           99.6\% &              - &           99.8\% \\ \hline
\end{tabular}
\caption{Results across all 9 settings for each county, along with cross-scenario comparisons. Performance is measured by taking the mean speed for the day across all vehicles, with the standard deviation shown in parentheses. Safety is represented by the maximum number of supervisors needed for that scenario throughout the day, with the mean shown in brackets. The best system performance as assessed by the maximum number of supervisors across AV types is in \textbf{bold} for each county-penetration rate scenario. The `gain' rows show the percent \textit{rise} in speed and percent \textit{reduction} in the maximum number of supervisors required for the CCAV setting relative to the UCAV setting, Baseline 1, and Baseline 2, respectively. The baselines are explained in the main text, \ch{along with further details and analysis of the information presented}.}
\label{table:results_all}
\end{table*}

\begin{figure}
    \centering
    \includegraphics[width=0.47\textwidth]{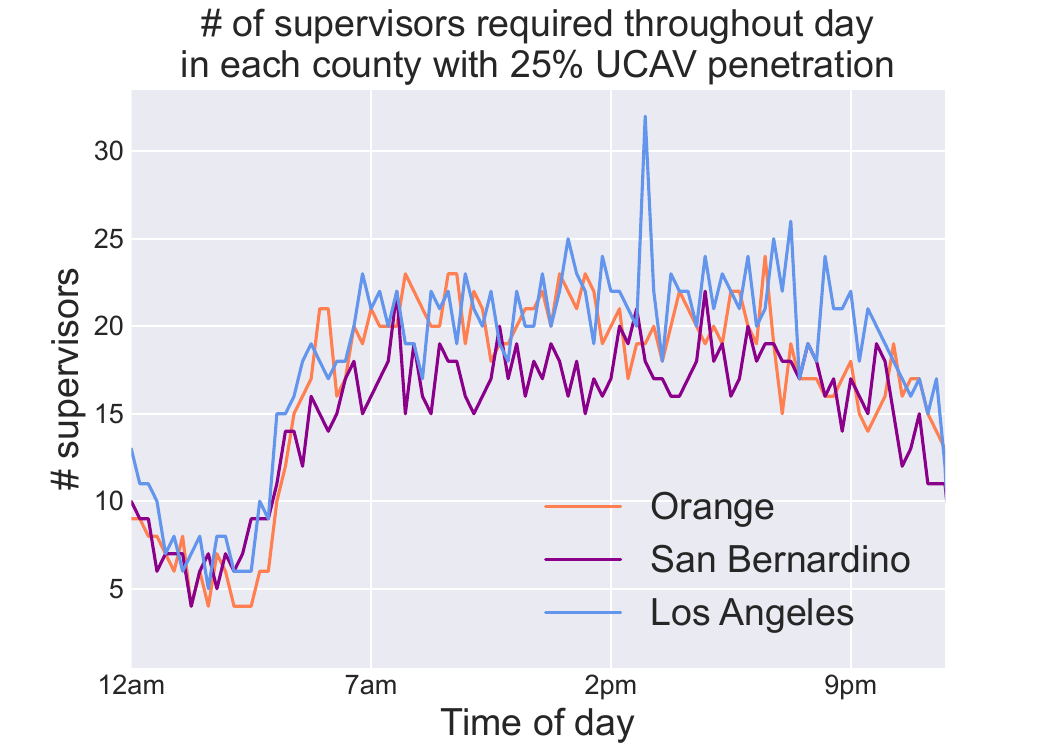}
    \caption{County supervision requirements with a 25\% UCAV penetration rate. The maximum number of supervisors required during each 15-minute interval is plotted. This can be unpredictable; note LA County's afternoon spike.}
    \label{fig:sup_requirements}
\end{figure}

\begin{figure}
    \centering
    \includegraphics[width=0.47\textwidth]{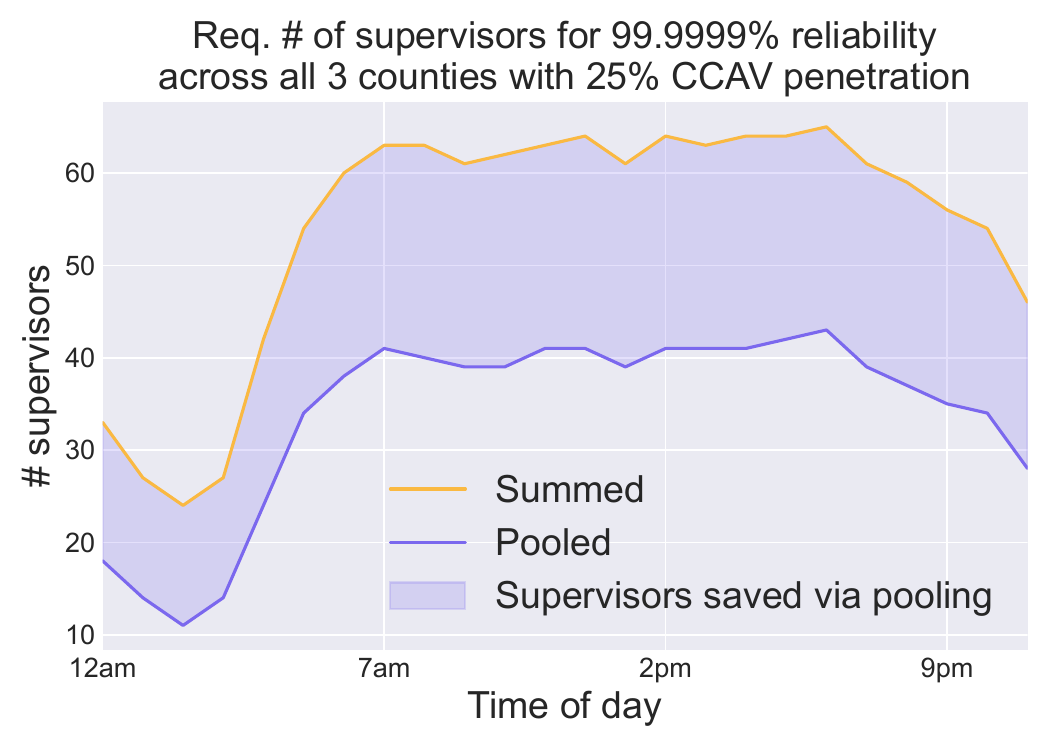}
    \caption{Supervision aggregation benefits: the gold line shows the sum total of supervisors needed across the three counties to achieve six `nines' of reliability when supervision occurs on a per-county basis. The purple line shows the number needed when supervision tasks are pooled across counties.}
    \label{fig:pooling_lineplot}
\end{figure}

\begin{figure}
    \centering
    \includegraphics[width=0.4\textwidth]{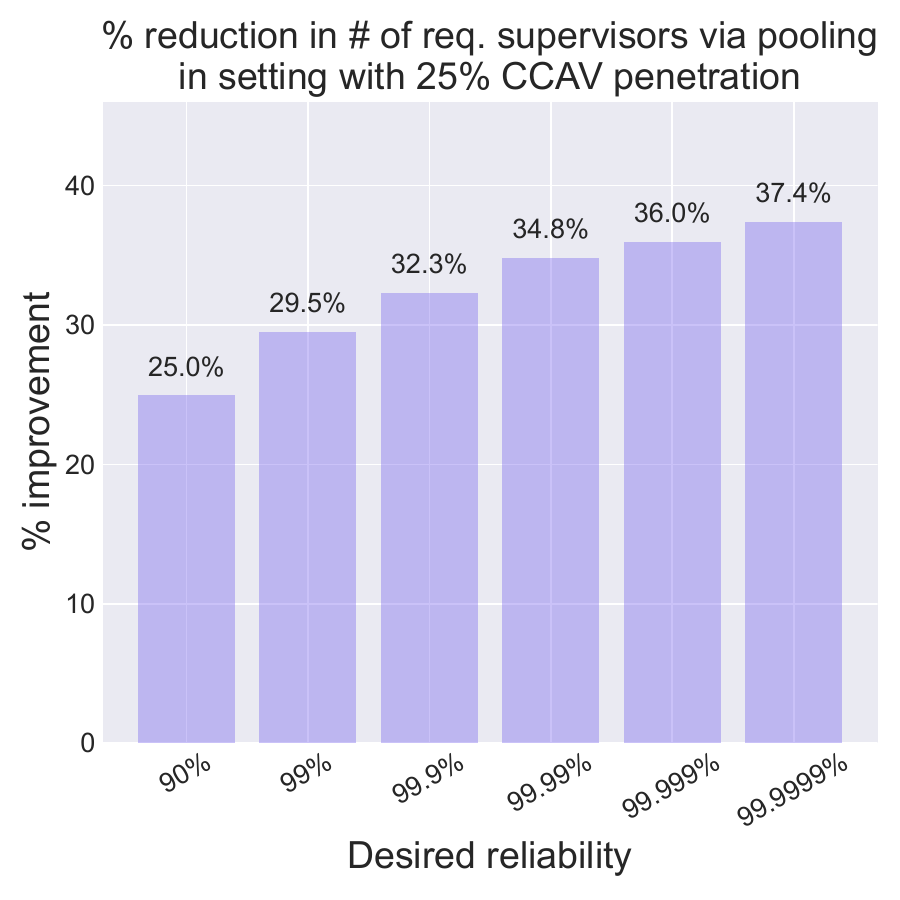}
    \caption{As reliability demands increase, so do the pooling benefits. \ch{The rightmost bar corresponds to the shaded area in~\Cref{fig:pooling_lineplot}.}}
    \label{fig:pooling_barplot}
\end{figure}

\begin{table}[ht]
    \centering
    \begin{tabular}{|c|c|c|c|}
        \hline
             UCAV/CCAV pen. rate & 25\% & 50\% &	75\% \\ \hline
             Cross-county `nines' &  &  & \\
             (orders-of-magnitude) of & 1 & 4 & 6 \\
             reliability gained via CCAVs & & & \\ \hline
             Avg. `nines' gain across all settings & \multicolumn{3}{c|}{3.67} \\
        \hline
    \end{tabular}
    \caption{Average orders-of-magnitude of system reliability improvements across counties for CCAVs relative to UCAVs. (E.g., 99.9997\% compared to 99.93\% is a gain of two `nines'.) The final cell shows the average benefit across all penetration rates.}
    \label{table:nines}
\end{table}

Results for the empirical number of supervisors required over the course of a day are shown in~\Cref{table:results_all}. For each of the 27 scenarios (a product of the three counties, three penetration rates, and three AV types), the table shows performance and safety values. Performance is assessed as the mean speed across vehicles averaged over the duration of the day. The associated standard deviation is shown in parentheses. Safety is measured as the maximum number $\hat{k}$ of supervisors needed to monitor AV merges over the 24 hours. The mean value is included in brackets to provide a sense of how many supervisors would be active on average. 

Note that the supervisor means are substantially lower than the maximums; this helps illustrate the variance in the supervision requirements (see \Cref{fig:sup_requirements}) and also provides some intuition on how busy the supervision team would be. Even as the present research leaves to future work questions related to accounting for broader human factors (buffer time between tasks, breaks, etc.), the maximum-mean gaps across the scenarios suggests that an appropriately-sized supervision team may not often operate at capacity, which could ease the task for the remote operators.

Given the range of possible AV deployment paradigms, we include two baselines. The first (`Baseline 1') represents a scenario in which the remote operators must supervise \ch{\textit{all AV merges}}. This highlights the benefit of reachability-based supervision, since an AV in Baseline 1 is supervised even when merging onto an empty highway. 
The second (`Baseline 2') represents a \ch{more demanding} case in which all AVs are supervised for the \ch{\textit{duration of their trip}} in a 1:1 human:AV ratio. This approximates the case in which an in-vehicle human is supervising --- as is common in many vehicles with advanced driver-assist technologies today --- except we allow the number of operators to `teleport' from a vehicle that left the system to a vehicle that just arrived. 
In reality, this baseline would require more supervisors (and the comparative gains of scalable supervision would be greater), since trips extend beyond the interstate system. This baseline emphasizes the benefit of remotely-located operators who are not tied to any single AV. Performance values do not change with the baselines and thus are excluded to avoid redundancy. 
The CCAVs' cooperative behavior often results in narrow differences for baseline supervisor values between the UCAV/NCAV setting and the CCAV setting \ch{(since cooperation can slightly alter trajectories, thus perturbing lane selections and trip durations)}. Where this occurred, these were averaged to ease comparison. 

The best performance for each of the nine county-penetration rate pairs (as assessed by the maximum number of supervisors needed throughout the day) is shown in bold. 
CCAVs performed best in seven of nine scenarios, with reductions up to 43\% in the number of supervisors needed relative to the UCAV scenario. Importantly, they did so without similar reductions in system performance. The system's average speed only dropped by a maximum of 3\% across the scenarios.
NCAVs performed well in San Bernardino. The authors hypothesize this may be due to the low population density relative to the other counties; cooperative behavior may be less necessary when highways are sparsely populated. One can observe how the UCAV-NCAV gap (which is due to NCAVs' connectivity that results in reduced uncertainty about their location) compares to the NCAV-CCAV gap (which is due to CCAVs' cooperative, merge-assisting driving policy) across scenarios.

\addtolength{\textheight}{-0.5cm}   

The combined fourfold benefits of the (1) remote, (2) reachability-based supervision scheme with vehicles that are (3) connected, and (4) cooperative are more evident in the CCAVs' improvements relative to the baselines. Compared to when all AV merges are supervised, the CCAV scenarios achieve 34-73\% reductions in the number of required supervisors. Relative to the baseline with 1:1 human:AV supervision, we find that CCAVs reduce the supervision requirements by more than 99\% across all scenarios. 

CCAVs also provide safety benefits when considering the minimum number of operators $k^*$ to achieve a long-term desired level of supervision reliability. To illustrate this, we find the greatest hourly empirical arrival rate $\hat{\lambda_l}$ and the associated $\hat{\mu_l}$ value for the UCAV setting in each county-penetration rate pair, as well as the corresponding values for each CCAV setting. We then compute and compare the $1 - \epsilon$ reliability value (where $P_{k^*} < \epsilon$) across both settings, letting $k^* = \hat{k}$, the maximum number of supervisors required over the course of the day in the \textit{CCAV setting only}. For example, in Los Angeles with a 50\% penetration rate, we compare the reliability achieved with the separate $\hat{\lambda_l}$ and $\hat{\mu_l}$ values found for UCAVs and CCAVs, but in both cases consider that only 36 supervisors total are available. Reliability is often assessed in orders-of-magnitude terms, where the emphasis is on how many `nines' of reliability a system can provide: a 99.93\% value corresponds to three nines of reliability, whereas a 99.9997\% value corresponds to five nines of reliability, and so forth.
\Cref{table:nines} shows the orders-of-magnitude gains in reliability averaged across counties, and then averaged again across penetration rates. The experiments show CCAVs achieve an average of 3.67 more nines than UCAVs in the scenarios studied \ch{due to the CCAVs' connectivity and cooperative behavior}.

Finally, we assess the benefits of aggregating supervisors across the adjacent counties. Such pooling is a known tool in queue theory to improve operational performance in some settings~\cite{cattani2005pooling}. In our case we compare the reliability levels achieved when a separate team of remote operators supervises each of the three counties (shown as `summed' in \Cref{fig:pooling_lineplot}) to those achieved when supervision tasks are aggregated across all the counties and thus one team may supervise them (shown as `pooled' in the figure). The plot demonstrates how such pooling can reduce supervision requirements. 

Interestingly, the benefits of pooling grow as reliability demands increase, as shown in \Cref{fig:pooling_barplot}. Each bar is proportional to the shaded portion in \Cref{fig:pooling_lineplot} for different desired reliability levels. Results in both figures are shown for the setting with a 25\% penetration rate of CCAVs since that would occur prior to the higher rates, but the qualitative behavior is similar across all scenarios analyzed. 

\section{Conclusion}

This work outlined a framework DISCES for assessing safety and performance in data-informed, realistic traffic settings, and applied it to AV merges in mixed-autonomy traffic. Our findings suggest remote, event-based human supervision may be a viable avenue towards enhancing safety in near-term AV adoption, and that cooperative driving strategies can provide significant reductions in supervisory burdens --- as well as related gains in system reliability --- without proportional drops in traffic system performance. The work also indicates that scalability can improve by pooling supervision tasking across wider areas. 

Future work could investigate more sophisticated CCAV driving strategies, consider data-driven alternatives to reachability-based supervision \ch{(such as predictive models)}, extend supervision to other challenging driving problems, or tackle the human factors questions involved. Additional research could also apply the DISCES framework to other safety-critical events in entirely new traffic settings.








\bibliographystyle{bib_files/IEEEtran}
\bibliography{bib_files/IEEEabrv,bib_files/bibliography}

\end{document}